\documentclass[twocolumn,showpacs,amsmath,amssymb,prl]{revtex4}
\usepackage{amsmath}
\usepackage{graphicx}
\usepackage{color}
\usepackage{bm}
\usepackage[colorlinks,citecolor=blue]{hyperref}

\begin{document}
\title{Quantum-criticality-induced strong Kerr nonlinearities in optomechanical systems}
\author{Xin-You L\"{u}$^{1}$}
\author{Wei-Min Zhang$^{1,4}$}
\author{Sahel Ashhab$^{1,2}$}
\author{Ying Wu$^{3}$}
\author{Franco Nori$^{1,2}$}
\address{$^1$Advanced Science Institute, RIKEN, Wako-shi, Saitama 351-0198, Japan\\$^2$Physics Department, The University of Michigan, Ann Arbor, Michigan 48109-1040, USA
\\$^3$Wuhan National Laboratory for Optoelectronics and School of Physics, Huazhong University of Science and Technology, Wuhan 430074, People's Republic of China
\\$^4$ Department of Physics, National Cheng Kung University, Tainan 70101, Taiwan}

\date{\today}

\begin{abstract}
We investigate a hybrid electro-optomechanical system that allows us to obtain controllable strong Kerr nonlinearities in the {\it weak-coupling regime}. We show that when the controllable electromechanical subsystem is close to its {\it quantum critical point}, strong photon-photon interactions can be generated by adjusting the intensity (or frequency) of the microwave driving field. Nonlinear optical phenomena, such as the appearance of the photon blockade and the generation of nonclassical states (e.g., Schr\"{o}dinger cat states), are predicted in the {\it weak-coupling regime}, which is feasible for most current optomechanical experiments.
\end{abstract}


\maketitle {\it Introduction.---} Strong optical nonlinearity, as one of the central issues of quantum optics, give rise to many strictly quantum effects, such as photon blockade \cite{1,1.5}, optical solitons \cite{3.1}, quantum phase transitions \cite{4,4.1}, quantum squeezing \cite{4.2} and optical switching with single photon \cite{5}. These nonlinear optical effects have been demonstrated in cavity QED systems, where the quantum coherence in the atom \cite{1,1.5} (or artificial atom \cite{6,6.1,6.2,6.3}) generates strong effective photon-nonlinearities.

Recently, cavity optomechanics has become a rapidly developing research field exploring nonlinear couplings via radiation pressure between the electromagnetic and mechanical systems \cite{7,7.1,7.2}. It has been shown {\it theoretically} that strong optical nonlinear effects (and relevant applications, such as generating nonclassical state, photon blockade, multiple sidebands, photon-phonon transistors, and optomechanical photon measurement) can be realized in single-mode \cite{7.3,8,9,9.1,10,10.1,10.2} or two-mode optomechanical systems (OMSs) \cite{11,12} in the single-photon strong-coupling regime, where the optomechanical coupling at the single-photon level $g_{a}$ exceeds the cavity decay rate $\kappa_{a}$ ($g_{a}>\kappa_{a}$). However, in most experiments to date \cite{12.2,12.3,12.4}, $g_{a}$ is much smaller than $\kappa_{a}$ ($g_{a}/\kappa_{a}\sim 10^{-3}$). Only a few new-type optomechanical setups, using ultracold atoms in optical resonators ($g_{a}/\kappa_{a}\sim10^{-1}$) \cite{13} or optomechanical crystals ($g_{a}/\kappa_{a}\sim10^{-2}$) \cite{14}, can approach the single-photon strong-coupling regime. On the other hand, a strong optical driving field may enhance the optomechanical coupling by a factor $\sqrt{n}$, where $n$ is the mean photon number in the cavity \cite{15,16,17}. But such enhancement comes at the cost of losing the nonlinear character of the photon-photon interaction.
\begin{figure}
\includegraphics[width=0.43\textwidth]{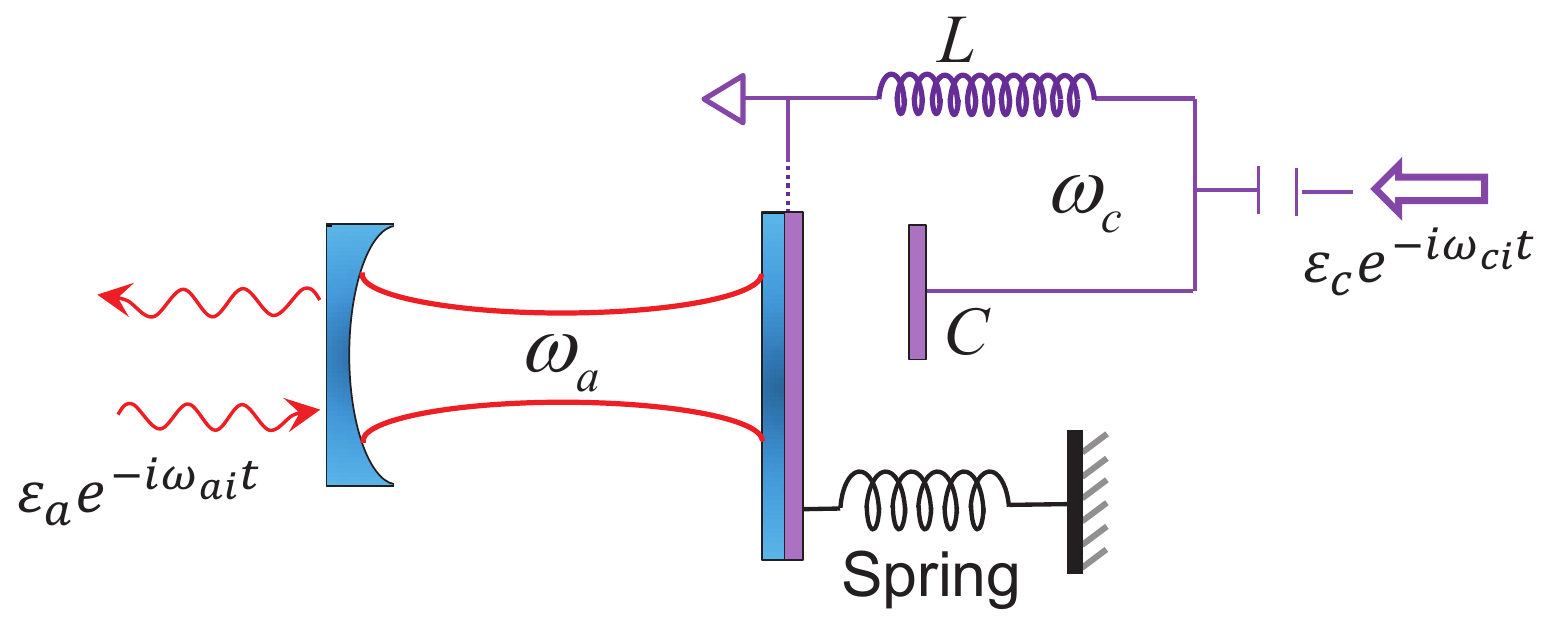}
\caption{(Color online) Schematic sketch of the hybrid electro-optomechanical system, where a mechanical oscillator couples to both an optical cavity and a microwave {\it LC} resonator.}\label{fig1}
\end{figure}

Given the above, it is highly desirable to find a new method for obtaining strong Kerr nonlinearities in OMSs in the {\it weak-coupling regime} ($g_{a}\ll\kappa_{a}$). In this Letter, we investigate the Kerr nonlinear effects of the optical field in a hybrid electro-optomechanical system containing a mechanical oscillator coupled to both an optical cavity and a microwave {\it LC} resonator (see Fig.\,\ref{fig1}) \cite{22,24,24.5}. We find that the eletromechanical subsystem (the mechanical oscillator and microwave resonator) displays quantum criticality. One can drive the electromechanical subsystem close to the quantum critical regime by applying a strong microwave field to the microwave resonator. The quantum criticality can induce a strong Kerr nonlinearity in the optical cavity, even if the optomechanical systems (the optical cavity and mechanical oscillator) is in the weak-coupling regime. This strong Kerr nonlinearity can be demonstrated by the existences of photon blockade and nonclassical states (e.g., Schr\"{o}dinger cat states) of the cavity field when the electronechanical subsystem approaches the quantum critical point. Furthermore, the strong Kerr nonlinearity can also be controlled easily by tuning the intensity (or frequency) of the microwave driving field. This provides a promising route for experimentally observing strong Kerr nonlinearities in OMSs in the weak-coupling regime.

\maketitle {\it Quantum criticality of the electromechanical subsystem.---} In the hybrid electro-optomechanical system of Fig.\,\ref{fig1}, the mechanical oscillator is parametrically coupled to both the optical cavity and the microwave resonator. The microwave resonator is driven by a strong field with amplitude $\varepsilon_{c}$ and frequency $\omega_{ci}$, where $\varepsilon_{c}$ is related to the input microwave power $P$ and microwave decay rate $\kappa_{c}$ by $|\varepsilon_{c}|=\sqrt{2P\kappa_{c}/\hbar\omega_{ci}}$. In the frame rotating with frequency $\omega_{ci}$, the Hamiltonian for the hybrid systems is written as \cite{24.6}

\begin{align}\label{eq1} 
\hat{H}/\hbar &=\delta_{c}\hat{c}^{\dagger}\hat{c}+\omega_{a}\hat{a}^{\dagger}\hat{a}+\omega_{b}\hat{b}^{\dagger}\hat{b}+g_{a}\hat{a}^{\dagger}\hat {a}\left( \hat{b}^{\dagger}+\hat{b}\right)\nonumber
\\
&\;\;\;\;+g_{c}\hat{c}^{\dagger}\hat {c}\left(\hat{b}^{\dagger}+\hat{b}\right)  +\varepsilon_{c}\left(\hat{c}^{\dagger }+\hat{c}\right),
\end{align}
where the detuning $\delta_{c}=\omega_{c}-\omega_{ci}$ and the microwave frequency $\omega_{c}=1/\sqrt{LC}$, $g_{a}$ ($g_{c}$) denotes the optomechanical (electromechanical) coupling strength at the single-photon level, and $\hat{a}$ ($\hat{b}$ or $\hat{c}$) is the annihilation operator of the optical cavity (the mechanical oscillator or the microwave resonator).  Taking a strong microwave driving field and following the standard linearization procedure (shifting $\hat{c}$ and $\hat{b}$ with their stable-state mean value $\alpha$ and $\beta$) \cite{25,26,27}, the Hamiltonian can be transformed into
\begin{align}\label{eq2} 
\hat{H}_{\rm opt}/\hbar&
=\Delta_{c}\hat{c}^{\dag}\hat{c}+\tilde{\omega}_{a}\hat{a}^{\dag}\hat{a}+\omega_{b}\hat{b}^{\dag}\hat{b}\nonumber
\\
&\;\;\;\;+g_{a}\hat{a}^{\dag}\hat{a}\left(\hat{b}^{\dagger}+\hat{b}\right)-G\left(\hat{c}^{\dag}+\hat{c}\right)\left(\hat{b}^{\dag}+\hat{b}\right),
\end{align}
where
$G=g_{c}\sqrt{\frac{\varepsilon^2_{c}}{\kappa^2_{c}+\Delta^2_{c}}}$ is the linearized electromechanical coupling strength; $\Delta_{c}=\delta_{c}-\frac{2g^2_{c}\varepsilon^2_{c}}{\omega_{b}(\kappa_{c}^2+\Delta^2_{c})}$ and $\tilde{\omega}_{a}=\omega_{a}-\frac{2g^2_{c}\varepsilon^2_{c}}{\omega_{b}(\kappa_{c}^2+\Delta^2_{c})}$ are, respectively, the effective microwave detuning and optical frequency including the radiation-pressure-induced optical resonance shift. Notice that $G$ and $\Delta_{c}$ can be easily controlled by tuning the power and frequency of the microwave driving field \cite{30}.
\begin{figure}
\includegraphics[width=0.5\textwidth]{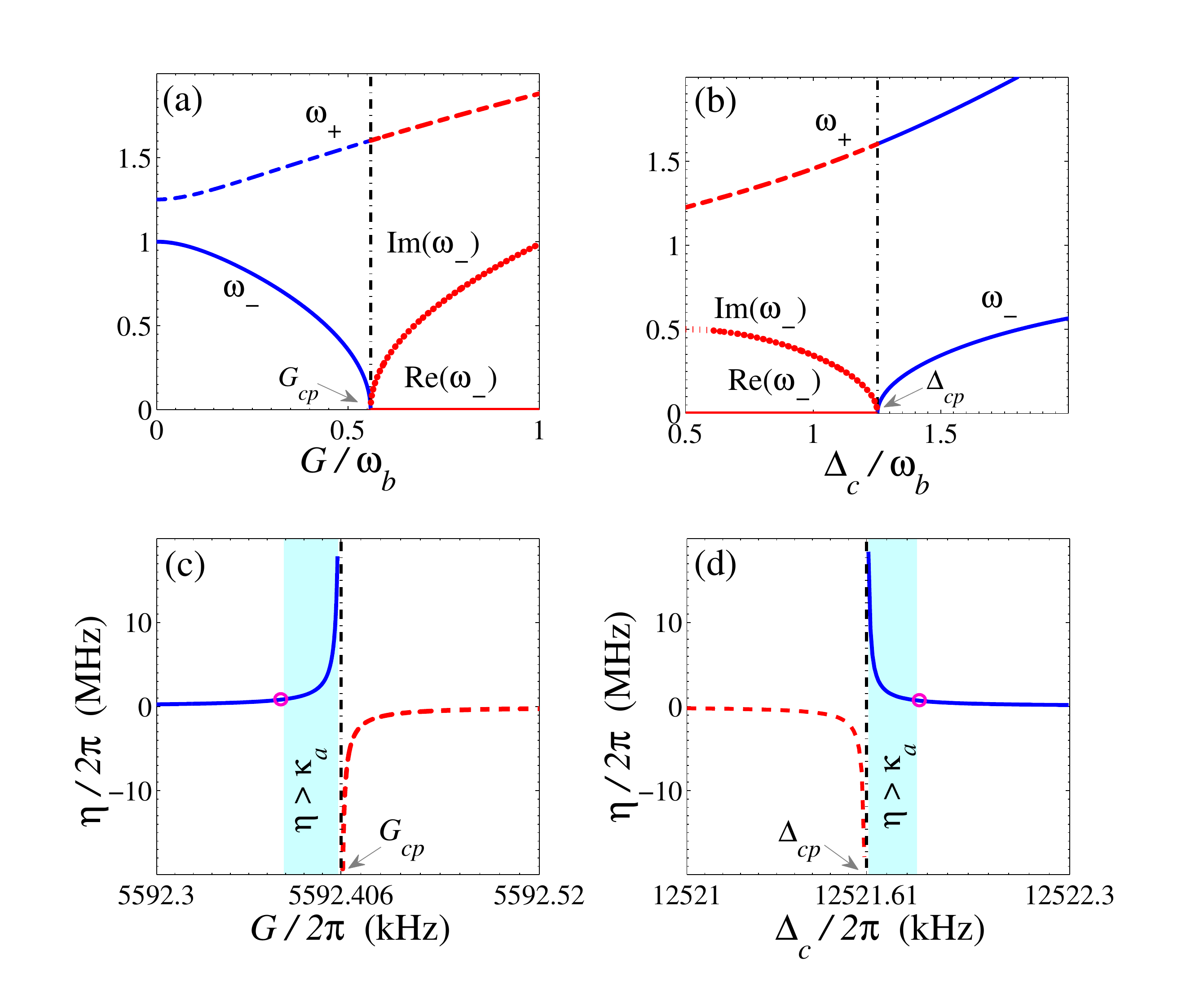}
\caption{(Color online) (a,b) Quantum criticality of the electromechanical subsystem, characterized by the normal-mode frequency $\omega_{\pm}/\omega_{b}$. (c,d) Strong Kerr-nonlinearity given by the photon-photon interaction strength $\eta$ in the optical cavity, as a function of the adjustable parameters $G$ and $\Delta_{c}$ controlled by the microwave driving field. The red circles and shaded area in (c,d) correspond, respectively, to the regimes $\eta=\kappa_{a}$ and $\eta>\kappa_{a}$. The black dot-dashed vertical lines indicate the quantum critical points $G_{cp}$ and $\Delta_{cp}$. Other system parameters are taken as: $\omega_{b}$/2$\pi$=10 MHz, $g_{a}/\omega_{b}$=$g_{c}/\omega_{b}$=$10^{-3}$, $\kappa_{a}/\omega_{b}=0.1$, 
$\kappa_{c}/\omega_{b}=0.127$, $\Delta_{c}/\omega_{b}=1.251$ (a,c), and $G/\omega_{b}=0.5595$ (b,d).}\label{fig2}
\end{figure}

To show quantum criticality in the electromechanical subsystem through the control of the microwave deriving field, we first diagonalize the electromechanical subsystem by a Bogoliubov transformation $\hat{R}=M\hat{B}$. Here, the canonical operators are $\hat{R}^{T}=(\hat{c},\hat{c}^{\dag},\hat{b},\hat{b}^{\dag})$ and $\hat{B}^{T}=(\hat{B}_{-},\hat{B}_{-}^{\dag},\hat{B}_{+},\hat{B}_{+}^{\dag})$,
and $M$ is the transformation matrix (the explicit form of $M$ is shown in the supplemental material \cite{30}). Then, the Hamiltonian $\hat{H}_{\rm opt}$ becomes
\begin{align}\label{eq3} 
\hat{H}_{\rm opt}/\hbar&=\omega_{-}\hat{B}_{-}^{\dag}\newline\hat{B}_{-}+\omega_{+}\hat{B}_{+}^{\dag}\newline\hat{B}_{+}+\tilde{\omega}
_{a}\hat{a}^{\dag}\hat{a} \nonumber\\
&-g_{-}
\hat{a}^{\dagger}\hat{a}\left(\hat{B}_{-}^{\dagger}\newline+\hat{B}_{-}\right)
+g_{+}\hat{a}^{\dagger}\hat{a}\left(\hat{B}_{+}^{\dag}\newline+\hat{B}_{+}\right),
\end{align}
where $\omega_{\pm}$ are the normal mode frequencies of the electromechanical subsystem, and $g_{\pm}$ are the effective coupling strengths between the optical photon and the normal modes,
\begin{subequations}
\label{eq4} 
\begin{align} 
&\omega_{\pm}^{2}=\frac{1}{2}\left(\Delta_{c}^{2}+\omega_{b}^{2}\pm
\sqrt{\left(\omega_{b}^{2}-\Delta_{c}^{2}\right)^{2}+16G^{2}\Delta_{c}\omega_{b}}\right)\\
&g_{\pm}=g_{a}\sqrt{\frac{\omega_{b}(1\pm {\rm cos 2\theta})}{2\omega_{\pm}}},\;\;\;\;{\rm tan}2\theta=\frac{4G_{c}\sqrt{\Delta_{c}\omega_{b}}}{\Delta^{2}_{c}-\omega^{2}_{b}}.
\end{align}
\end{subequations}
Equation\,(\ref{eq4}) shows that $\omega^{2}_{-}$ becomes zero (negative) when $G=G_{cp}=\sqrt{\Delta_{c}\omega_{b}}/2$ ($G>G_{cp}$), as shown in Fig.\,\ref{fig2}{\color{red} {(c)}}. This corresponds to a quantum criticality \cite{31}, namely, the normal mode $\omega_{-}$ will change from a standard harmonic oscillator ($G<G_{cp}$) to a free particle, and further becomes dynamically unstable ($G>G_{cp}$) as $G$ crosses its critical value $G_{cp}$, as shown in Fig.\,\ref{fig2}{\color{red} (a)}.  Physically, when $G$ approaches (or exceeds) $G_{cp}$, the effective potential of the normal mode $\omega_{-}$ becomes increasingly flat (or inverted). Since $G$ can be easily varied by tuning the power and frequency of the microwave driving field, this quantum criticality can be easily realized in experiments.

{\it Quantum-criticality-induced strong Kerr nonlinearities.---}
We find that when the electromechanical subsystem approaches its quantum critical region, the optical cavity shows a strong Kerr nonlinearity. To show this quantum-criticality-induced strong Kerr nonlinearity, the Hamiltonian $\hat{H}_{\rm opt}$ should be further diagonalized in a displaced-oscillator representation, $\hat{H}_{\rm opt}\rightarrow \hat{V}^{\dag}\hat{H}_{\rm opt}\hat{V}$, where $\hat{V}=e^{{\hat{\mathcal{P}}\hat{a}^{\dagger}\hat{a}}}$ and $\hat{\mathcal{P}}=\zeta_{-}\hat{\mathcal{P}}_{-}-\zeta_{+}\hat{\mathcal{P}}_{+}$ with $\hat{\mathcal{P}}_{j}=\hat{B}^{\dagger}_{j}-\hat{B}_{j}$ $(j=\pm)$, $\zeta_{\pm}=g_{\pm}/\omega_{\pm}$. The result is 
 \begin{align}\label{eq5} 
\frac{\hat{H}_{\rm opt}}{\hbar} & =\tilde{\omega}_{a}\hat{a}^{\dag}\hat{a}-\eta\
\hat{a}^{\dag}\hat{a}\hat{a}^{\dag}\hat{a}+\omega_{-}\hat{B}_{-}^{\dag}\newline\hat{B}_{-}
+\omega_{+}\hat{B}_{+}^{\dag}\newline\hat{B}_{+}.
\end{align}
Here the photon-photon interaction strength
\begin{align}\label{eq6} 
\eta=\frac{g^{2}_{a}}{\omega_{b}-4G^{2}/\Delta_{c}}.
\end{align}
It can be seen in Figs.\,\ref{fig2}{\color{red} (c,d)} that even in the {\it weak-coupling regime} $g_{m}\ll\kappa_{m}$ ($m=a,c$), a large photon-photon interaction $\eta$ ($\eta>\kappa_{a}$) can still be obtained when $G$ (or $\Delta_{c}$) is in the quantum critical regime. In particular, Fig.\,\ref{fig2} shows that when the coupling strength $G$ (or the detuning $\Delta_{c}$) is close to its quantum critical point, a very small normal mode frequency $\omega_{-}$ is obtained, which induces a large photon-photon interaction with $\eta\propto1/\omega_{-}$. The interesting ranges of $G$ and $\Delta_{c}$ are respectively on the order of $0.1$ kHz and $1$ kHz for the quantum critical region $\eta>\kappa_{a}$ (shaded area in Fig.\,\ref{fig2}), and this parameter precision is experimentally realizable \cite{32}.  

More importantly, the obtained large photon-photon interaction directly characterizes a strong optical Kerr nonlinearity. This is because, as we will show later, the quantum criticality also significantly suppresses the sideband phonon transitions in the optomechanical subsystem. Thus, the quantum-criticality-induced strong self-Kerr nonlinearity is very different from previous investigations in the usual OMSs, where the strong self-Kerr nonlinearity is reachable only in the single-photon strong-coupling ($g_{a}>\kappa_{a}$) and the resolved sideband ($\kappa_{a}\ll\omega_{b}$) regimes \cite{8,11,12}. To demonstrate the strong Kerr nonlinearity in the present system, we should calculate the steady-state second-order correlation function of the optical field $g^{(2)}(0)$, and show explicitly the photon blockade effect [$g^{(2)}(0)\rightarrow0$] in the weak-coupling regime, as can be experimentally detected by a Hanbury-Brown-Twiss Interferometer \cite{1.5}. We will also calculate the dynamical evolution of the cavity field and show the periodic generation of noclassical states, which are experimentally detectable via quantum state tomography. Notice that the photon blockade \cite{8,11} and nonclassical states \cite{7.3}, as evidences of strong Kerr nonlinearities, were obtained in the OMSs only in the single-photon strong-coupling regime and the resolved sideband regime. 
\begin{figure}
\includegraphics[width=0.45\textwidth]{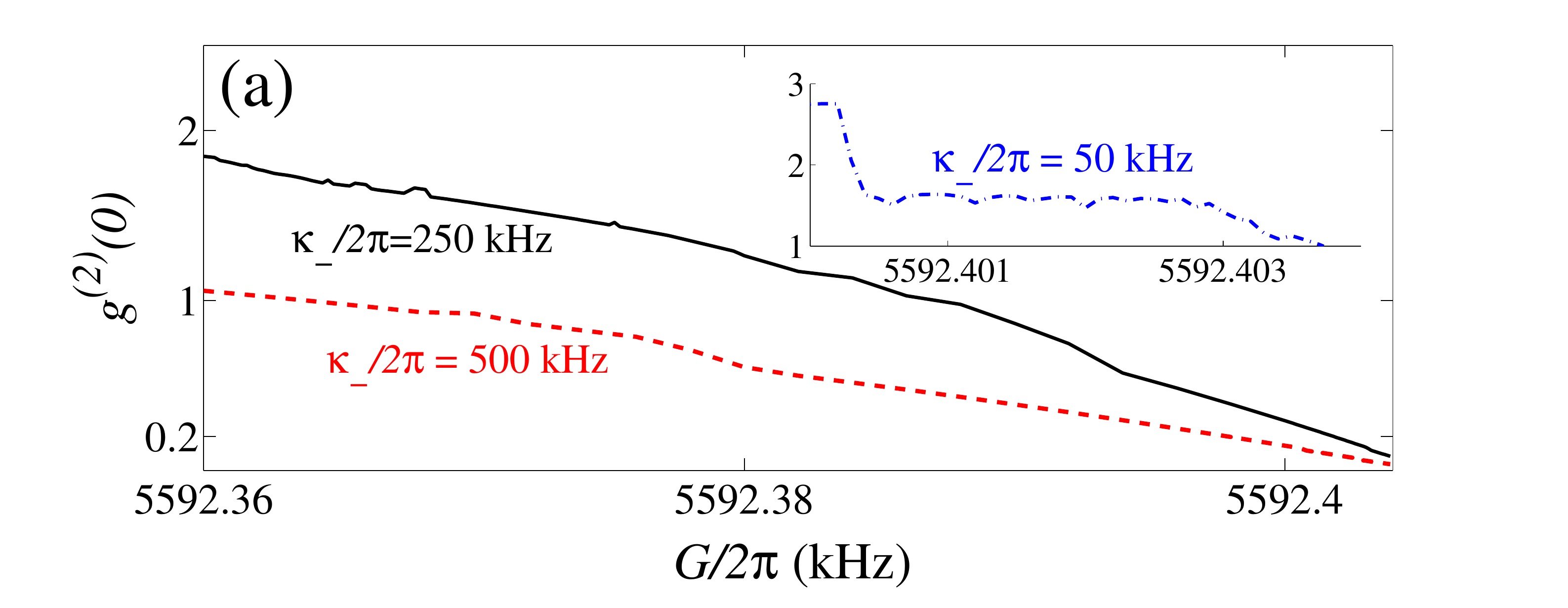}
\includegraphics[width=0.45\textwidth]{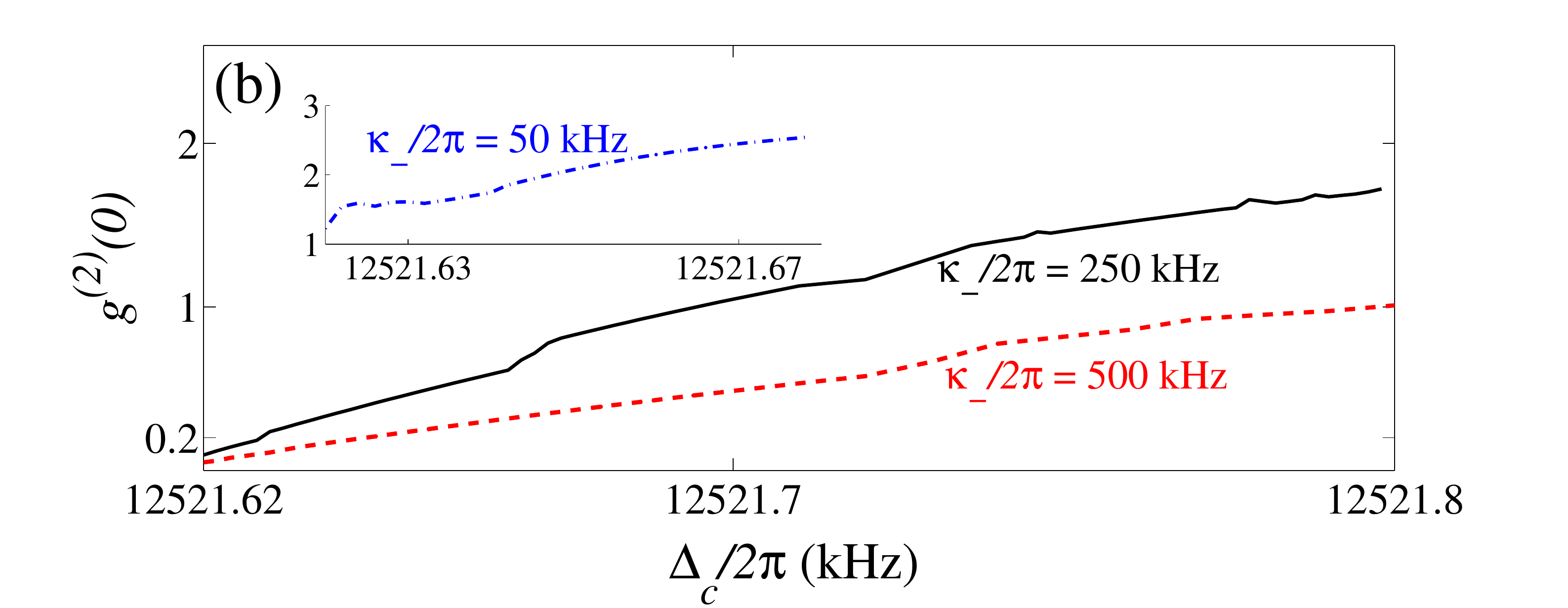}
\caption{(Color online) Equal-time second-order correlation function $g^{(2)}(0)$ versus: (a) coupling strength $G$, and (b) detuning
$\Delta_{c}$, for the decay rates $\kappa_{-}/2\pi=(500,250,50)$ kHz (corresponding to $\kappa_{c}/2\pi=(1270,620,110)$ kHz and 
$\kappa_{b}/2\pi=1$ kHz). The system parameters are the same as in Fig.\,\ref{fig2} except for $\Delta_{a}=\eta$, $\kappa_{+}/2\pi=500$ kHz.}\label{fig3}
\end{figure}

\maketitle {\it Photon blockade.---} We now drive the optical cavity with a weak laser field of frequency $\omega_{ai}$ and amplitude $\varepsilon_{a}$. The Hamiltonian of the system becomes
\begin{align}\label{eq7} 
\hat{H}_{\rm opt}'/\hbar & =\Delta_{a}\hat{a}^{\dag}\hat{a}-\eta\hat{a}^{\dag}\hat{a}\hat{a}^{\dag}\hat{a}+\varepsilon_{a}\left(\hat{a}^{\dagger}e^{-\hat{\mathcal{P}}}+e^{\hat{\mathcal{P}}}\hat{a}\right)\nonumber
\\
&+\omega_{-}\hat{B}_{-}^{\dag}\newline\hat{B}_{-}+\omega_{+}\hat{B}_{+}^{\dag}\newline\hat{B}_{+},
\end{align}
where all the similarity transformations used before have been taken into account, and $\Delta_{a}=\tilde{\omega}_{a}-\omega_{ai}$. Also, we may include the damping effect arising from the coupling of the optical field to the electromagnetic vacuum modes of the environment. Then, the dissipative dynamics of cavity mode $\hat{a}$ can be described by the quantum Langevin equation,
\begin{align}\label{eq8} 
\frac{\partial}{\partial
t}\hat{a}(t)=\frac{i}{\hbar}\left[\hat{H}_{\rm opt}',\hat{a}(t)\right]-\kappa_{a}\hat{a}(t)-\sqrt{2\kappa_{a}}e^{-\hat{\mathcal{P}}}\hat{f}_{\rm in}(t).
\end{align}
Here $\kappa_{a}$ is the decay rate of cavity mode $\hat{a}$ and $\hat{f}_{\rm in}$ is a vacuum noise operator satisfying $\langle\hat{f}_{\rm in}\hat{f}_{\rm in}^{\dagger}\rangle=\delta(t-t^{'})$, $\langle\hat{f}^{\dagger}_{\rm in}\hat{f}_{\rm in}\rangle=0$.

For a weak optical driving field, the quantum Langevin equations can be solved by truncating them to the lowest relevant order in $\varepsilon_{a}$ \cite{8}. The resulting two-photon correlation is given by $g^{(2)}(0)={\rm lim}_{t\rightarrow\infty}\langle\hat{a}^{\dagger}\hat{a}^{\dagger}\hat{a}\hat{a}\rangle(t)/\langle\hat{a}^{\dagger}\hat{a}\rangle^{2}(t)$ with
\begin{subequations}
\label{eq9} 
\begin{align}
&{\rm lim}_{t\rightarrow\infty}\langle\hat{a}^{\dagger2}\hat{a}^{2}\rangle(t)=\frac{2\varepsilon^{4}_{a}}{\kappa_{a}} {\rm Re}\!\int^{\infty}_{0}\!\!\!\!\!\!d\tau_{1}\int^{\infty}_{0}\!\!\!\!\!\!d\tau_{2}\int^{\infty}_{0}\!\!\!\!\!\!d\tau_{3}\nonumber
\\
&\times e^{2(-i\tilde{\Delta}_{a}+i\eta-\kappa_{a})\tau_{1}}e^{(i\tilde{\Delta}_{a}-\kappa_{a})\tau_{2}}e^{(-i\tilde{\Delta}_{a}-\kappa_{a})\tau_{3}}e^{-\Phi_{4}},\\
&{\rm lim}_{t\rightarrow\infty}\langle\hat{a}^{\dagger}\hat{a}\rangle(t)=\frac{\varepsilon^{2}_{a}}{\kappa_{a}}{\rm Re}\int^{\infty}_{0}\!\!\!\!\!d\tau
\,e^{-(i\tilde{\Delta}_{a}+\kappa_{a})\tau}e^{-\Phi_{2}},
\end{align}
\end{subequations}
where $e^{-\Phi_{4}}=\langle e^{\hat{\mathcal{P}}(\tau_{1}-\tau_{2})}e^{\hat{\mathcal{P}}(\tau_{1})}e^{-\hat{\mathcal{P}}(0)}e^{-\hat{\mathcal{P}}(-\tau_{3})}\rangle$, $e^{-\Phi_{2}}=\langle e^{\hat{\mathcal{P}}(\tau)}e^{-\hat{\mathcal{P}}(0)}\rangle$, and $\tilde{\Delta}_{a}=\Delta_{a}-\eta$. Note that 
$\hat{\mathcal{P}}=\zeta_{-}\hat{\mathcal{P}}_{-}-\zeta_{+}\hat{\mathcal{P}}_{+}$ is a complex operator including the microwave field $\hat{c}$ and the mechanical mode $\hat{b}$. The dynamics of $\hat{\mathcal{P}}_{j}(t)$ $(j=\pm)$ is given by $\hat{\mathcal{P}}_{j}(t)=e^{-\frac{i}{\hbar} (\hat{H}_{\rm opt}'-i\kappa_{j}/2)t}\hat{\mathcal{P}}_{j}(0)e^{\frac{i}{\hbar} (\hat{H}_{\rm opt}'+i\kappa_{j}/2)t}$, where the $\kappa_{j}$ are the effective decay rates of the electromechanical normal modes (we have also assumed a white vacuum noise on the microwave cavity and a thermal white noise bath coupling to the mechanical oscillator, so that the effective decay rates $\kappa_{j}$ are proportional to the original decay rates of the microwave resonator and the mechanical oscillator \cite{30}).

Assuming the microwave (mechanical) mode is initially in the coherent state $|\alpha\rangle$ ($|\beta\rangle$), and the optical field in the vacuum state, then the two-point correlation function $e^{-\Phi_{2}}$ and the four-point correlation function $e^{-\Phi_{4}}$ can be calculated \cite{30}. We numerically integrate Eqs.\,(\ref{eq9}) and show the dependence of $g^{(2)}(0)$ on both $G$ and $\Delta_{c}$ for different  decay rates $\kappa_{-}$ in Fig.\,\ref{fig3}, while the effective decay rate $\kappa_{+}$ is a fixed value due to its negligible effect on $g^{(2)}(0)$. Fig.\,\ref{fig3} shows that the photon blockade [$g^{(2)}(0)\rightarrow0$] occurs when the tunable parameter $G$ (or $\Delta_{c}$) approaches its quantum critical value even if the optomechanical coupling $g_{a}$ is very weak.

Furthermore, we find that the photon antibunching effect [$g^{(2)}(0)<1$] disappears when $\kappa_{-}\ll\kappa_{a}$ (see the insets in Fig.\,\ref{fig3}). The physical meaning of this result can be explained as follows. In the hybrid OMS, a relatively large decay rate $\kappa_{-}$ ($\kappa_{-}\sim\kappa_{a}$) with respect to the effective mechanical mode $\omega_{-}$ occurs when the electromechanical subsystem approaches the quantum critical point. This decay will significantly suppress the steady-state sideband transition in the electromechanical subsystem. Meanwhile, the very small $\omega_{-}$ near the quantum critical point effectively enhances the photon-photon interaction to $\eta>\kappa_{a}$ because $\eta\propto1/\omega_{-}$. Thus, the photon blockade can still be obtained in our system even if $\omega_{-}<\kappa_{a}$. However, for the usual OMSs, when the frequency of the mechanical oscillator is smaller than the decay rate of the cavity mode (out of the resolved sideband regime), the photon blockade will disappear due to the strong phonon sideband transition \cite{8,11}.
\begin{figure}
\includegraphics[width=0.23\textwidth]{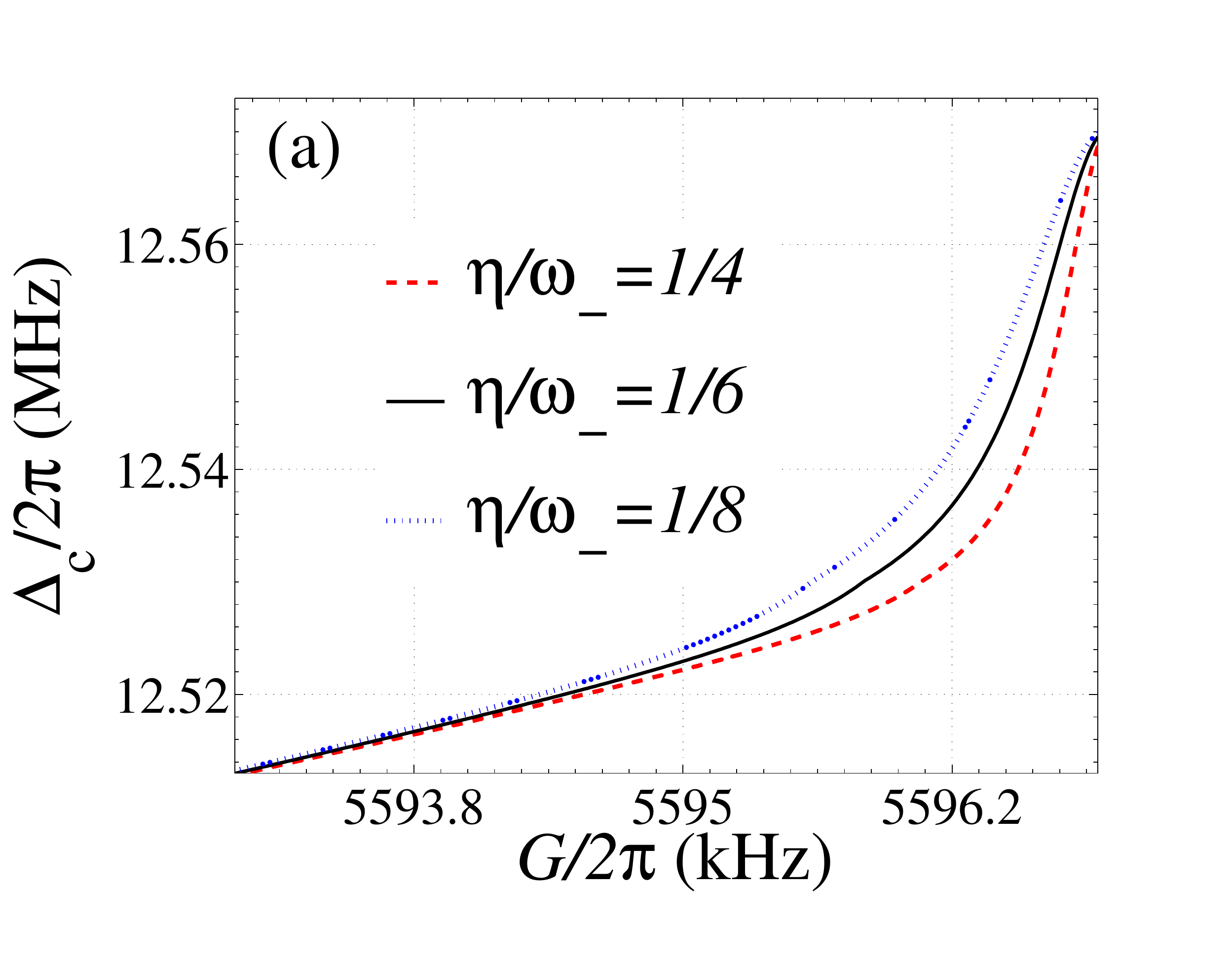}\includegraphics[width=0.23\textwidth]{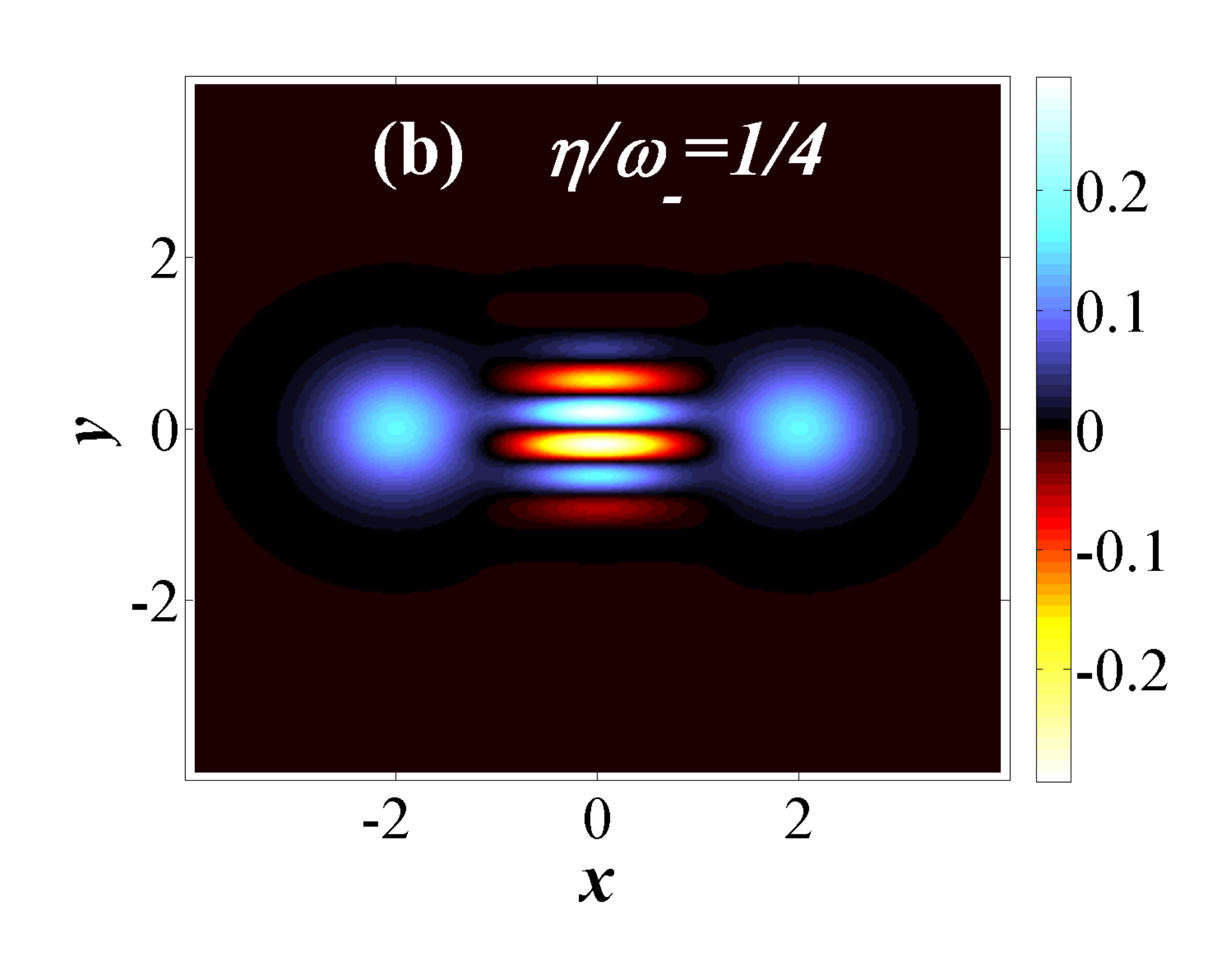}
\includegraphics[width=0.23\textwidth]{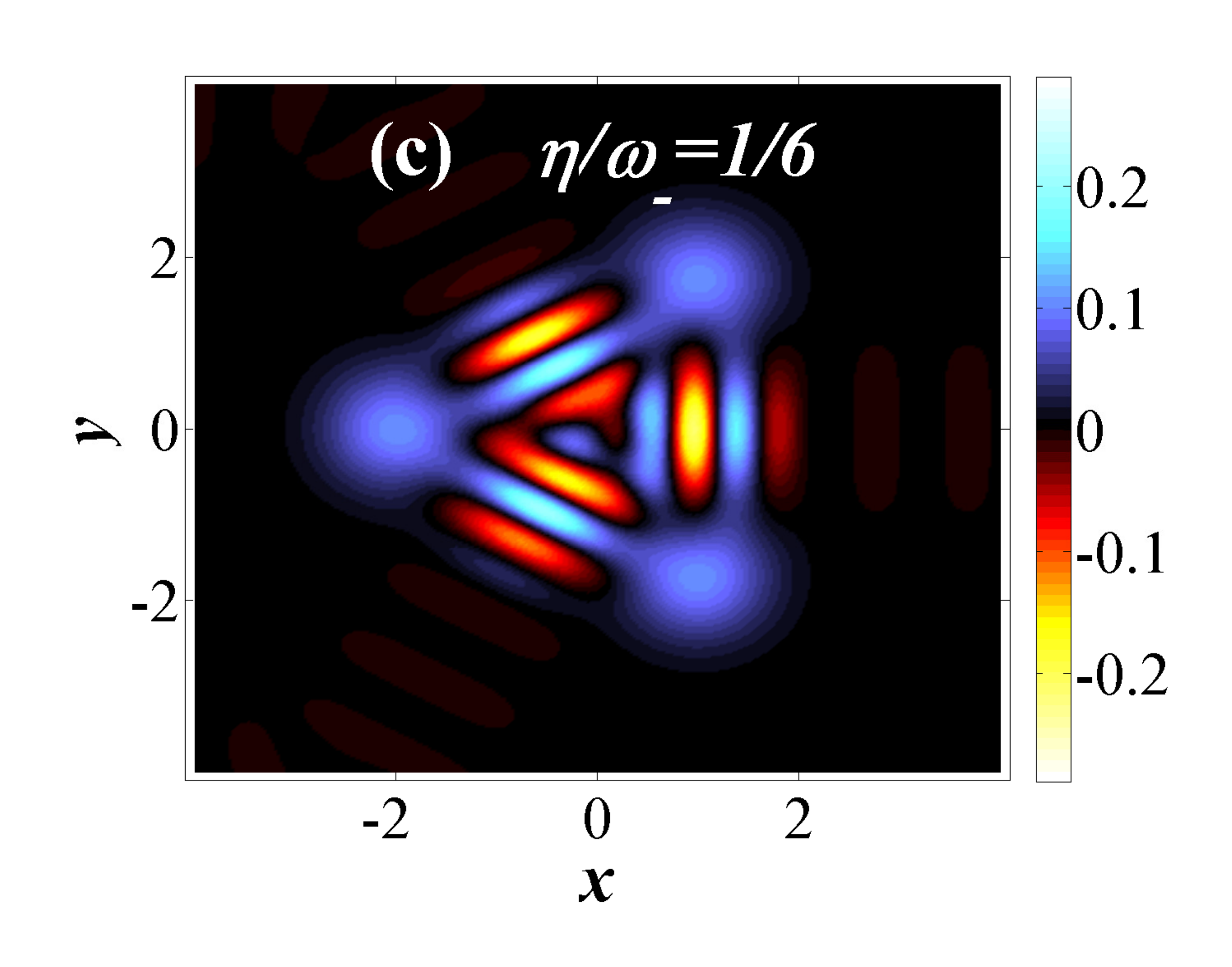}\includegraphics[width=0.23\textwidth]{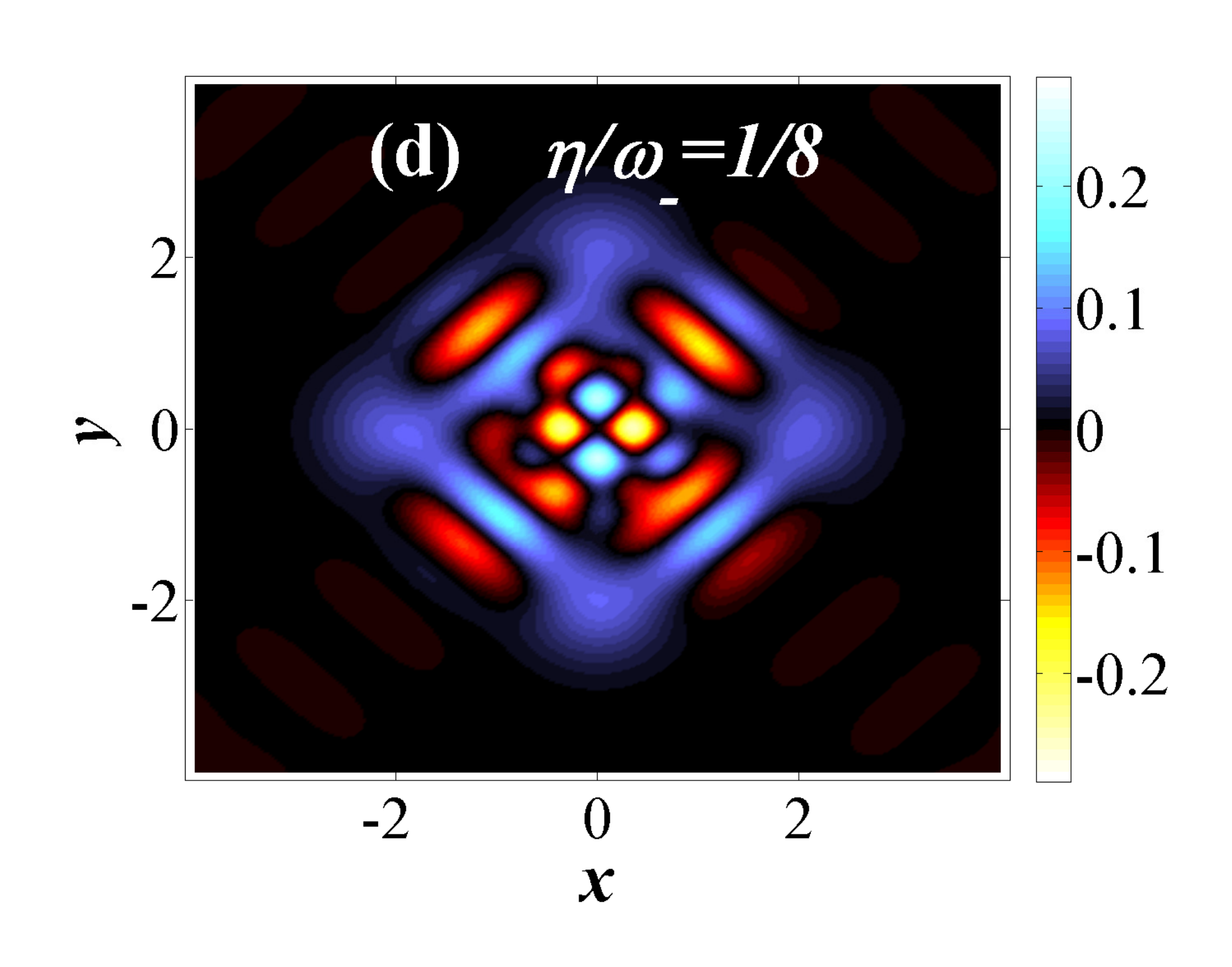}
\caption{(Color online) Parameter regimes (a) for obtaining the two- (b), three- (c) and four-component (d) Schr\"odinger cat state. The quadratures variables are $x=(\hat{a}+\hat{a}^{\dagger})/\sqrt{2}$, $y=-i(\hat{a}-\hat{a}^{\dagger})/\sqrt{2}$. The system parameters are the same as in Fig.\,\ref{fig2} except for $\Upsilon=2$.}\label{fig4}
\end{figure}

\maketitle {\it Nonclassical states.---} As demonstrated in previous studies \cite{7.3}, strong Kerr nonlinearities generally lead to the periodic in time generation of a broad variety of nonclassical states, (i.e., cat states) of the cavity field. With the help of the Hamiltonian (\ref{eq5}), we can obtain the time evolution operator in the interaction picture,
\begin{align}\label{eq10} 
&\hat{U}(t)=\hat{V}e^{-i(\hat{H}_{\rm opt}-\tilde{\omega}_{a}\hat{a}^{\dagger}\hat{a})t}\hat{V}^{\dagger}\nonumber
\\
&\approx e^{i\eta
\hat{a}^{\dagger}\hat{a}\hat{a}^{\dagger}\hat{a}t}\left\{e^{\zeta_{-}\hat{a}^{\dagger}\hat{a}\left[\hat{B}^{\dagger}_{-}(1-e^{-i\omega_{-}t})-\hat{B}_{-}(1-e^{i\omega_{-}t})\right]}\right\},
\end{align}
where the term corresponding to $\zeta_{+}$ has been omitted due to its negligible effect on the evolution of the cavity mode $\hat{a}$ ($\zeta_{+}/\omega_{b}\sim10^{-4}$) near the quantum critical point. If the cavity field $\hat{a}$ is initially in a coherent state $|\Upsilon\rangle$, the cavity field at time $t_{n}$, $t_{n}=2n\pi/\omega_{-}$ $(n=1,2...)$ will evolve into the state
\begin{align}\label{11}
|\Psi_{a}(t_{n})\rangle=e^{-|\Upsilon|^2/2}\sum^{\infty}_{m=0}\frac{\Upsilon^{m}}{\sqrt{m!}}e^{i\frac{2n\pi\eta}{\omega_{-}}m^2}|m\rangle_{a}.
\end{align}
The state $|\Psi_{a}(t_{n})\rangle$ is a multi-component cat state, depending on the value of $\eta/\omega_{-}$. Fig.\,\ref{fig4} shows the different multi-component cat states for different values of $\eta/\omega_{-}$ near the quantum critical point. Figs.\,\ref{fig4}{\color{red} (b,c,d)} present the specific realization of two-, three- and four-component cat states, respectively. Here we should point out that the system's damping (given by $\kappa_{a},\kappa_{c},\kappa_{b}$) has been ignored. In principle, this approximation is valid when the cut-off time $t_{n}\ll1/\kappa_{a},1/\kappa_{c},1/\kappa_{b}$. The above result indicates that the quantum-criticality-induced strong Kerr nonlinearities in this hybrid OMS can generate nonclassical states by cutting off the optomechanical interaction at the appropriate time, which can be detected via Wigner tomography.  

{\it Conclusion.}--In summary, we have identified a mechanism for obtaining strong Kerr nonlinear effects in a hybrid OMS in the {\it weak-coupling regime}. The photon-photon interaction is controllable through the adjustable parameters $G$ and $\Delta_{c}$, and the sideband phonon transitions can be suppressed. The photon blockade and nonclassical states are demonstrated near the quantum critical point. This may provide a new avenue for experimentally realizing strong optical nonlinearities in the {\it weak-coupling regime} and largely enrich the parameter scope for implementing quantum information processing and quantum metrology with cavity OMSs.      

{\it Acknowledgements---}
XYL thanks Jing Liu, Jie-Qiao Liao, Guang-Ri Jing, Jing Zhang and Wei Cui for valuable discussions. FN was partially supported by the ARO, JSPS-RFBR contract No. 12-02-92100, Grant-in-Aid for Scientific Research (S), MEXT Kakenhi on Quantum Cybernetics, and the JSPS via its FIRST program.
YW is was partially supported by the National Science Foundation (NSF) of China (Grants No. 10975054), the National Fundamental Research Program of China (Grant No. 2012CB922103). WMZ is supported by the National Science Council of ROC under Contract No. NSC-99-2112-M-006-008-MY3, and the National Center for Theoretical Science of Taiwan. XYL was supported by Japanese Society for the Promotion of Science (JSPS) Foreign Postdoctoral
Fellowship No. P12204 and the NSF of China under grant number 11005057.

\end{document}